# Fabrication of yttrium–iron–garnet/Pt multilayers for the longitudinal spin Seebeck effect

Tatsuhiro Nozue[1,a)], Takashi Kikkawa[2,3,b)], Tomoki Watamura[2], Tomohiko Niizeki[3], Rafael Ramos[3], Eiji Saitoh[2,3,4,5,6], and Hirohiko Murakami[1]

[1]*Future Technology Research Laboratory, ULVAC, Inc., Tsukuba 300-2635, Japan*
[2]*Institute for Materials Research, Tohoku University, Sendai 980-8577, Japan*
[3]*Advanced Institute for Materials Research, Tohoku University, Sendai 980-8577, Japan*
[4]*Department of Applied Physics, The University of Tokyo, Tokyo 113-8656, Japan*
[5]*Center for Spintronics Research Network, Tohoku University, Sendai 980-8577, Japan*
[6]*Advanced Science Research Center, Japan Atomic Energy Agency, Tokai 319-1195, Japan*

(ABSTRACT)

For longitudinal spin Seebeck effect (LSSE) devices, a multilayer structure comprising ferromagnetic and nonmagnetic layers is expected to improve their thermoelectric power. In this study, we developed the fabrication method for alternately stacked yttrium–iron–garnet (YIG)/Pt multilayer films on a gadolinium gallium garnet (GGG) (110) substrate, GGG/[YIG(49 nm)/Pt(4 nm)]$_n$ ($n$ = 1–5) based on room-temperature sputtering and *ex-situ* post-annealing method and we evaluated their structural and LSSE properties. The fabricated [YIG/Pt]$_n$ samples show flat YIG/Pt interfaces and almost identical saturation magnetization $M_s$, although they contain polycrystalline YIG layers on Pt layers as well as single-crystalline YIG layers on GGG. In the samples, we observed clear LSSE signals and found that the LSSE thermoelectric power factor (PF) increases monotonically with increasing $n$; the PF of the [YIG/Pt]$_5$ sample is enhanced by a factor of ~28 compared to that of [YIG/Pt]$_1$. This work may provide a guideline for developing future multilayer-based LSSE devices.

[a] tatsuhiro_nozue@ulvac.com

[b] t.kikkawa@imr.tohoku.ac.jp

Recent advances in the miniaturization of computing and mobile devices have increased the need for small power generators. In this context, the longitudinal spin Seebeck effect (LSSE) has gained attention, since it enables thermoelectric power generation with a small and simple structure consisting of a bilayer of ferromagnetic material (FM) and nonmagnetic metal (NM).[1–4] In an LSSE, when a temperature gradient is applied in the bilayer, a magnon flow in the FM is converted into a spin current in the NM by the interfacial exchange interaction. Subsequently the spin current in the NM generates a thermoelectric voltage via the inverse spin Hall effect, if the spin–orbit coupling in the NM is strong enough (e.g., Pt).[5–9]

Theoretically, high thermoelectric efficiency is expected from LSSE devices, since their output power is not limited by the Wiedemann–Franz law for conventional thermoelectric elements based on the (charge) Seebeck effect.[2,4,10] However, in practice, the LSSE thermoelectric power is very small.[4] This has driven a search for materials to improve the LSSE power.[2,4] Besides, the structural design of thermoelectric elements is also a research target to increase LSSE power.[2,4] Recently, LSSE power was found to be enhanced in alternately stacked FM/NM multilayers.[4,11–15] Ramos *et al.* reported $[Fe_3O_4/Pt]_n$ multilayer LSSE devices formed by repeated growth of $n$ ferromagnetic oxide $Fe_3O_4$ and NM Pt bilayers, using pulsed laser deposition (PLD) and sputtering, respectively.[12,13] They found that the voltage as well as the power generated by the LSSE increase with increasing $n$. They attributed the LSSE voltage enhancement to the increase in spin-current amplitude in the multilayers, which also enhances the LSSE power in combination with the increase in electrical conduction with $n$, because of the parallel connection of the conductive Pt layers.[4,12] A similar experimental result was reported also in $[CoFeB/Pt]_n$ multilayers.[11] Hence, the design of the FM/NM multilayers has received attention in terms of fundamental physics as well as applications.[4]

In this letter, we report the development of a fabrication method for high-quality [YIG/Pt]$_n$ multilayers, formed by $n$ times repeated growth of YIG/Pt bilayers based on sputtering and post-annealing, and show their structural and magnetic properties as well as the LSSE results. Yttrium–iron–garnet (YIG: $Y_3Fe_5O_{12}$) is a ferrimagnetic insulator exhibiting an exceptionally long magnon-diffusion length and is the fundamental material for LSSE and magnonics research.[4,16] In addition, because of its highly insulating property, the use of YIG in LSSE devices allows us to avoid a possible overlap with the anomalous Nernst effect of the FM layer, which may have been an issue in previous ferromagnetic metals $Fe_3O_4$- and CoFeB-based LSSE devices.[11-13] [YIG/Pt]$_n$ multilayers are fascinating for future LSSE-based thermoelectric application and other spintronic devices based on, for instance, magnon-mediated current drag and magnon valve effects.[17–21] However, there is no report focusing on its fabrication, the reason of which may be due to the difficulty in stacking YIG/Pt bilayers with maintaining clean and flat interfaces. In general, for crystallizing YIG layers, high-temperature annealing (at ~ 800°C) is necessary.[17] However, there are two fundamental problems in this high temperature process: (1) Pt films on YIG layers are easily deformed and may become aggregations and (2) the constituents such as oxygen tend to be easily desorbed from YIG layers for long-time high-temperature annealing. To obtain high quality [YIG/Pt]$_n$ multilayers by overcoming these issues, we developed the method based on *in-situ n*-times YIG/Pt repetition sputtering and one-time rapid-thermal-annealing (RTA) process with a face-to-face proximity configuration. Here, each Pt (YIG) layer becomes robust against the deformation/aggregation (desorption), since each Pt (YIG) is sandwiched by YIG (Pt) layers and also the total annealing time is reduced by the RTA process, giving the solution to the above problems. Our versatile [YIG/Pt]$_n$ fabrication method reported here will expand the scope of application of [YIG/Pt]$_n$ spintronic devices and LSSE-based thermoelectric power generators.

The multilayer samples were deposited on 10×10×0.5 mm$^3$ Gd$_3$Ga$_5$O$_{12}$ (GGG) (110) substrates by a dc- and rf-magnetron sputtering system, QAM-4 STS (ULVAC Kyushu), at room temperature (RT). The QAM-4 is equipped with four on-axis sputtering cathodes with 50-mm-diameter targets in a high-vacuum chamber. Before sputtering, GGG substrates were annealed in a furnace at 900 ºC in air for 30 min, in a face-to-face proximity configuration. Then, the YIG and Pt films were sputtered alternately and sequentially without breaking the vacuum to form GGG/[YIG/Pt]$_n$ with clean interfaces. Here, YIG films were deposited at an rf-power of 150 W using a sputtering gas of Ar + 2 vol.% O$_2$ with a pressure of 0.10 Pa, at a deposition rate of 0.26 nm min$^{-1}$. Pt films were deposited at a dc-power of 20 W and Ar pressure of 0.10 Pa, at a rate of 1.14 nm min$^{-1}$.

Since the as-deposited YIG films were amorphous and nonmagnetic, post-annealing was necessary to crystallize the YIG layers. A thin YIG layer (~2 nm) was deposited on the top Pt layer as a protective layer to prevent deformation of the top Pt film during the post-annealing. The *ex-situ* RTA in air was performed, where the multilayered samples were heated up to 825 ºC by taking 60 s and were kept for only 200 s, during which the YIG films became crystallized. In the RTA process, we adopted the face-to-face proximity configuration, in which the surfaces of the top YIG layers (~2 nm) of two identical GGG/[YIG/Pt]$_n$/cap-YIG samples are faced with each other to avoid possible desorption of the constituents, such as Y and Fe, from their surfaces.

For LSSE measurements, to apply a temperature difference, $\Delta T$, the sample was sandwiched between two AlN plates. The lower AlN plate, which was in contact with the bottom of the GGG substrate, was heated using a Peltier module, while the temperature of the upper AlN plate, in contact with the top surface of [YIG/Pt]$_n$ (with a contact area of $5 \times 2$ mm$^2$), was kept at RT by connecting a heat bath. To measure $\Delta T$, thermocouples were attached to the upper and bottom AlN plates.[22] We note that, although the estimated $\Delta T$ value may contain error due mainly to interfacial thermal resistance between the sample and AlN plates,[23-25] its variation between each measurement was confirmed to be negligibly small (only 1.2% variation in the SSE voltage $V_{SSE}$ for 6-times different measurements using the same sample). This situation allows us to compare the LSSE coefficient $S = (V_{SSE}/L_{sample})/(\Delta T/D_{sample})$[4] between the [YIG/Pt]$_n$ samples with different $n$ number [$L_{sample}$ = 5 mm (Pt length), $D_{sample}$ = 0.5 mm (GGG thickness)]. With the application of a constant $\Delta T$ of 3 or 6 K, the voltage, $V_{SSE}$, generated in the film was recorded as a function of the magnetic field $H$ applied along the [1−10] direction of the GGG substrate.

First, we briefly evaluated the surface morphology of GGG, GGG/YIG, and GGG/YIG/Pt/thin-YIG. Figure 1(a) shows the $1 \times 1$ μm$^2$ atomic force microscope (AFM) image of the annealed GGG (110) substrate. The surface exhibits a step-and-terrace structure with a root-mean-squared (RMS) roughness of 0.13 nm. We found that the surface of YIG deposited on GGG (110) after the RTA at 825 ºC exhibits a similar step-and-terrace structure, with an RMS of 0.10 nm [see Fig. 1(b)]. The result shows that YIG fabrication by RT sputtering and RTA can produce an extremely flat YIG film, comparable to the GGG substrates. Figure 1(c) shows an AFM image of the GGG/YIG/Pt/thin-YIG. Although no step-and-terrace structure was confirmed, its RMS roughness was 0.44 nm, which is smooth enough to grow further [YIG/Pt] multilayer films.

By X-ray reflection (XRR) measurements (not shown), we determined the average thickness of the YIG, Pt, and YIG protection layers to be $49.2\pm0.5$ nm, $4.00\pm0.08$ nm, and ~2.2 nm, respectively, so that the present sample structure can be written as GGG-subst./[YIG(49 nm)/Pt(4 nm)]$_n$/YIG(2 nm), $n$ = 1–5. By XRR, the averaged values of interfacial roughness were evaluated to be 0.29, 0.30, and 0.33 nm for $n$ =1, 2, and 3, respectively.

Figures 1(d)–(g) display the transmission electron microscope (TEM) images of a cross section of $n$ = 3 viewed along the [1–10] direction. The overall image shown in Fig. 1(d) reveals that each layer has a smooth surface/interface and is grown without macroscopic defects. The high-resolution (HR) image around the first-YIG/GGG interface [Fig. 1(e)] shows that the YIG layer is single crystalline and grown epitaxially on GGG. In contrast, the HR images around the first-YIG/first-Pt/second-YIG [Fig. 1(f)] and third-YIG/third-Pt [Fig. 1(g)] interfaces show that the YIG interlayers on Pt layers are grown as polycrystals [see the oblique lines in the HR images of the second- and third-YIG films, which are different from the first YIG (110) film]. The result could be attributed to the absence of seed GGG crystals for the YIG interlayers, unlike the first single-crystalline YIG layer; the as-deposited amorphous YIG layers on Pt are transformed, during the RTA, into polycrystalline YIG having many single-crystal grains with various crystalline

orientations. From the HR TEM images shown in Figs. 1(f) and 1(g), we confirmed that striped patterns exist in the Pt layers parallel to the interfaces. The distance between each stripe is ~0.23 nm, comparable to the (111) plane of Pt. This result suggests that, although the Pt layers are polycrystalline, the <111> axis is oriented almost perpendicular to the plane.

To obtain further information on the interfaces and element profile, we performed HR HAADF–STEM (or Z-contrast) observation. Figure 1(h) shows the Z-contrast image around the YIG/GGG interface of $n$ = 3. We found that the atomic arrangements of the first YIG agree with the <110> projection of the YIG lattice and exhibit the characteristic pattern of alternating Y and Fe atoms.[26] In Fig. 1(i), we show the Z-contrast image around the first-YIG/first-Pt/second-YIG interfaces. Small but finite regions of intermediate brightness were observed between the Pt and YIG layers (within ~1 nm, slightly greater than the interfacial roughness). From STEM–EDX line-scan measurements, we found that such regions contain mixed Pt, Y, Fe, and O elements, which could be attributed to possible inter-diffusion due to the high-temperature annealing process.[20] From STEM–EDX measurements, we also evaluated the chemical composition of the YIG layers and found an off-stoichiometric behavior of Y : Fe : O = 3 : 4.3 : 12, similar to the sputtered YIG films in Ref. 27.

Figure 2 shows the X-ray diffraction (XRD) $2\theta – \theta$ plots of the $n$ = 1, 3, and 5 samples. For all three samples, a diffraction peak was observed at 40.85º, slightly lower than that of GGG (440) diffraction, and its intensity (270 cps) was almost identical for all the samples. We assign the peak to the (440) diffraction from the first-YIG layer epitaxially grown on GGG, as confirmed in the TEM images. For $n$ = 3 and 5, we also observed other peaks at 28.73º (for $n$ = 5) and 32.22º (for $n$ = 3 and 5), which can be assigned to the YIG (400) and (420) diffractions, respectively, and may originate from the polycrystalline YIG interlayers. For all the samples a broad peak with Laue oscillations was observed at around 39.7º and can be attributed to the Pt (111) diffraction, as indicated in the TEM images [Figs. 1(f) and 1(g)].

From the peak position of YIG (440) diffraction, the out-of-plane lattice constant of the (first) single-crystalline YIG layers was determined to be 1.2486 nm, larger than that of both GGG (1.2383 nm) and bulk YIG (1.2376 nm). These results are consistent with previous reports on the epitaxial YIG thin films on GGG by PLD and sputtering,[26,28–33] and can be attributed to the off-stoichiometry of YIG.[29,32]

Subsequently, we investigated the magnetization, $M$, and ferromagnetic resonance (FMR) spectrum to reveal the magnetic properties of the YIG layers at RT. Figure 3(a) shows the $M$–$H$ curve of samples $n$ = 1 to 5. The intensity of $M$ was found to be proportional to the number of layers, which means that the saturation magnetization per volume, $M_S$, is almost equal for all the layers ($M_S$ = 108.00 ± 0.34 emu cm$^{-3}$) and that the $M$ magnitude of each YIG layer takes almost the same value. The $M_S$ value is approximately 20 % smaller than that of bulk YIG (140 emu cm$^{-3}$), which may be caused by the deficiency of Fe in our YIG films. In fact, an $M_S$ value of 103 emu cm$^{-3}$ is also reported, via FMR, for the single-crystalline YIG films sputtered on GGG with the chemical composition of Y : Fe = 3 : 4.4 (Fe deficiency).[27] Figure 3(b) shows the in-plane FMR derivative absorption spectra of $n$ = 1, 2, and 3, measured with microwaves of 1 mW at 9.45 GHz. For $n$ = 1, sharp FMR absorption was observed at 2290 Oe with a peak-to-peak linewidth, $\Delta H$, of 4.25 Oe. This linewidth is comparable to the reported values for the epitaxial YIG films prepared by PLD and sputtering.[26–28,30,32–34] For $n = 2$ and 3, at the same $H$, we also observed sharp absorptions with values of $\Delta H$ of 4.62 and 3.97 Oe, respectively, which can be attributed to the FMR absorption of their first single-crystalline YIG layers, and indicates that the first YIG layers show similar magnetic damping irrespective of the repetition-growth number $n$. In addition, for $n = 2$ and 3, we observed weak and broad absorptions at 2660 Oe with a $\Delta H$ of more than 50 Oe. The appearance of such peaks can be interpreted as the absorption of the polycrystalline YIG interlayers on Pt layers. The broadening of the absorption peak may originate from an enhanced magnetic damping due to grain-boundary scattering for the polycrystalline YIG layers, which is absent from the first single-crystalline YIG layers.

Finally, we investigated the LSSE for the [YIG/Pt]$_n$ multilayer. Before the LSSE measurement, we measured the resistance of the samples by the 4-probe method and found that the sheet resistance, $R_S$, of samples $n = 1$ to 5 is almost inversely proportional to $n$ ($R_S = 50.8, 26.0, 17.0, 13.7,$ and $10.5$ $\Omega$ for $n = 1, 2, 3, 4,$ and 5, respectively). This means that the resistivity of each Pt layer is almost identical. Figure 4(a) shows the LSSE results for $n = 1$ to 5. By applying $\Delta T$, clear LSSE signals were observed for all the samples and the LSSE curves were found to display hysteresis loops similar to the $M-H$ curve [see also Fig. 3(a)]. Although the $M$ intensity is proportional to $n$, the saturated values of the LSSE coefficient, $S_S$, exhibit a different behavior against $n$. The $S_S$ value of $n = 1$ [$S_S(1)$] is 0.18 $\mu$VK$^{-1}$. $S_S(2) = 0.38$ μVK$^{-1}$ for $n = 2$, being 210% as large as that of $S_S(1)$ [see also Fig. 4(b)]. For more stacked multilayers, $S_S(3) = 0.37$ and $S_S(4) = 0.38$ $\mu$VK$^{-1}$, giving almost the same value as that of $S_S(2)$. The $S_S$ value for $n = 5$ is $S_S(5) = 0.43$ $\mu$VK$^{-1}$, slightly larger than those for $n = 3$ and 4, and $S_S(5)/S_S(1) = 238\%$ [see also Fig. 4(b)].

The observed enhancement and almost saturated behavior of $S_S$ when $n \geq 2$ is different from the previous report in [Fe$_3$O$_4$/Pt]$_n$ multilayers.[12] In Ref. 12, $S_S(n)/S_S(1)$ increases to 255% at $n = 2$, then it monotonically increases with increasing $n$, and it finally takes the maximum value of 370% at $n = 6$, greater than our present value. In Ref. 12, to interpret the LSSE enhancement, the authors theoretically propose that the interfacial spin-current continuity between Fe$_3$O$_4$ and Pt leads to an increased amplitude of spin current in [Fe$_3$O$_4$/Pt]$_n$ multilayers, which is *more than twice* higher than that in a Fe$_3$O$_4$/Pt bilayer. On the other hand, our results can be explained simply by the double transfer of the spin currents into the Pt layer from the upper and lower YIG layers. The multilayer [YIG/Pt]$_n$ for $n \geq 2$ has $(n-1)$ sets of the sandwich structure of YIG/Pt/YIG. In such Pt layers, the spin currents are transferred from YIGs through the upper and lower interfaces. We here infer that the longitudinal magnon spin-current has the same magnitude in all the YIG layers, being motivated by the fact that (i) each YIG layer has an almost identical saturation magnetization and (ii) the robustness of spin current against YIG crystallinity is reported for YIG/Pt systems through LSSE measurements.[35] As a result, the spin-current amplitude of Pt in YIG/Pt/YIG takes the value twice as large as that in YIG/Pt. In this scenario, the open circuit voltage for [YIG/Pt]$_n$ may change by a factor of $(2-1/n)$ compared to $n = 1$, resulting in an increase in the LSSE coefficient $S_S$ with $n$ and giving the maximum $S_S$ value twice larger that of $n = 1$ for $n \gg 1$. In Fig. 4(b), we compare the $S_S$ values calculated by the above model (dashed line) with those obtained experimentally. We found reasonable agreement between them, although the experimental results show a slightly rapid increase with *n*. Therefore, the mechanism proposed here may play an important role in the present LSSE enhancement and saturation. We attribute the discrepancy between the previous [$Fe_3O_4$/Pt]$_n$ and present [YIG/Pt]$_n$ results to the difference in the samples' interfacial condition; the interfaces of the [$Fe_3O_4$/Pt]$_n$ multilayers are highly epitaxial, while those of the [YIG/Pt]$_n$ are *not* epitaxial. This may cause an increased loss of spin current at the YIG/Pt interfaces than at the $Fe_3O_4$/Pt interfaces and may make the previously-assumed spin-current continuity condition in Ref. 12 unreasonable for the present [YIG/Pt]$_n$ systems.

Although the LSSE coefficient is almost saturated when $n \geq 2$, the resistance of the samples decreases monotonically with increasing *n*, meaning that the power factor (PF) still increases with *n*. In Fig. 4(c) we plot the PF, defined as $S_S^2/R_S$.[4] The PF value of $n = 5$ is 0.02 pWK$^{-2}$, which is 27.5 times greater than that of $n = 1$. The results show that the [YIG/Pt]$_n$ multilayers are beneficial in terms of PF.

In conclusion, we fabricated multilayers consisting of GGG/[YIG(49 nm)/Pt(4 nm)]$_n$/YIG(2 nm) ($n = 1$ to 5) and investigated their structural and magnetic properties and their LSSE against the number of layers *n*. We show that well-crystallized [YIG/Pt]$_n$ with flat interfaces can be grown by RT sputtering and subsequent *ex-situ* face-to-face RTA processes. Using these samples we observed clear LSSE signals and found that PF increases monotonically with increasing *n*, although the LSSE coefficient *S* is almost saturated when $n \geq 2$. The knowledge about the [YIG/Pt]$_n$ fabrication reported here may be important for industrial applications of LSSE multilayer elements.

(Acknowledgments)

This work was supported by ERATO "Spin Quantum Rectification Project" (JPMJER1402) from JST, Grant-in-Aid for Scientific Research on Innovative Area "Nano Spin Conversion Science" (JP26103005) from JSPS KAKENHI, and ULVAC, Inc.

(Figures)

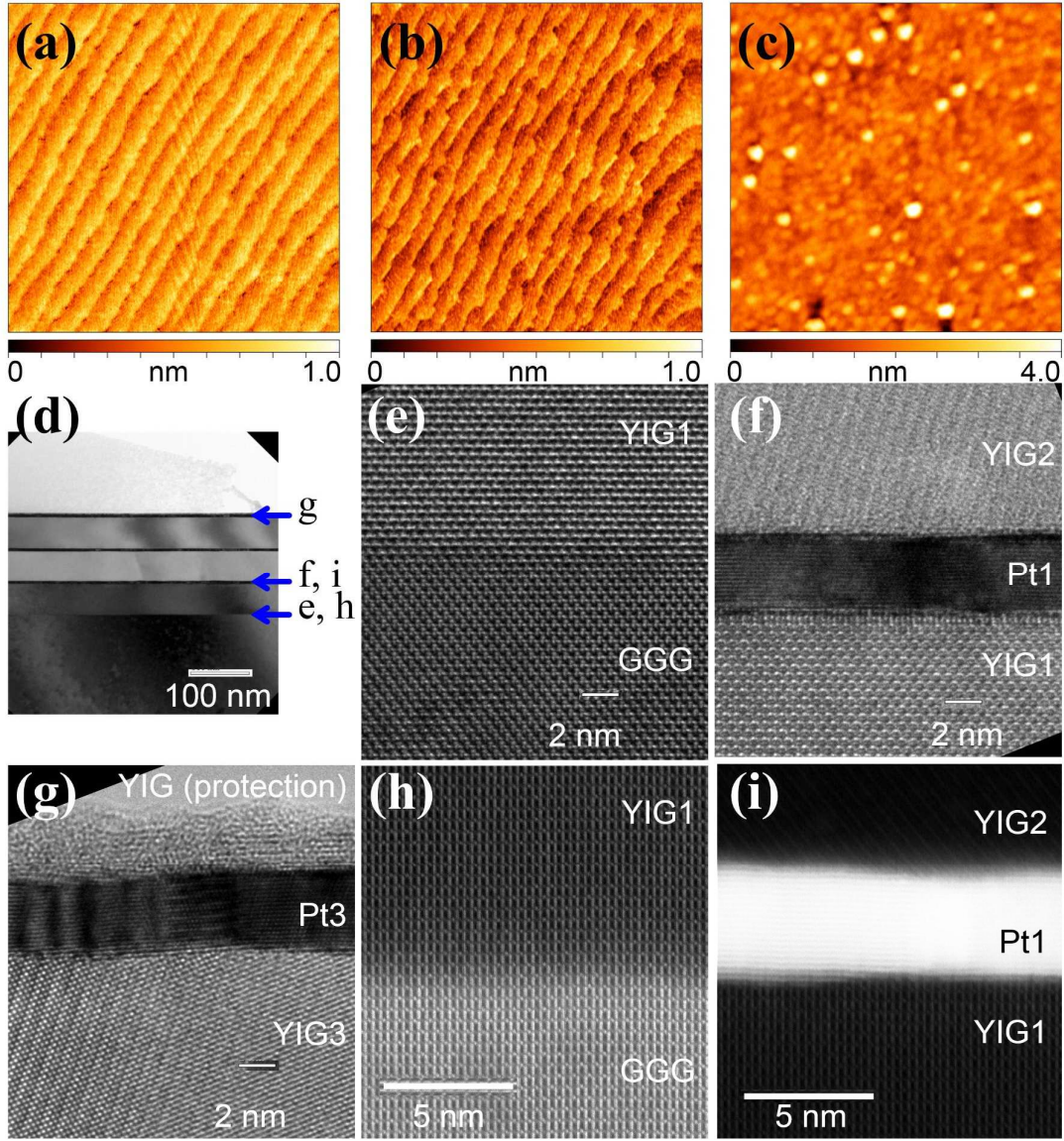

FIG. 1. (a)–(c) 1 μm × 1 μm AFM images of surfaces of the (a) GGG (110) substrate after annealing at 900 ºC, (b) YIG film deposited on GGG substrate after RTA at 825 ºC, and (c) YIG protection film deposited on GGG-subst./YIG/Pt(5 nm) after RTA at 825 ºC. (d)–(i) TEM and HAADF–STEM images of GGG/[YIG(49 nm)/Pt(4 nm)]$_3$/YIG(2 nm), ($n$ = 3). (d) An overall TEM image. (e), (f), (g) HR TEM images of the interfaces indicated by the arrows labeled as (e), (f), and (g) in (d). (h) – (i) HR HAADF–STEM images of the interfaces indicated by the arrows labeled as (h) and (i) in (d).

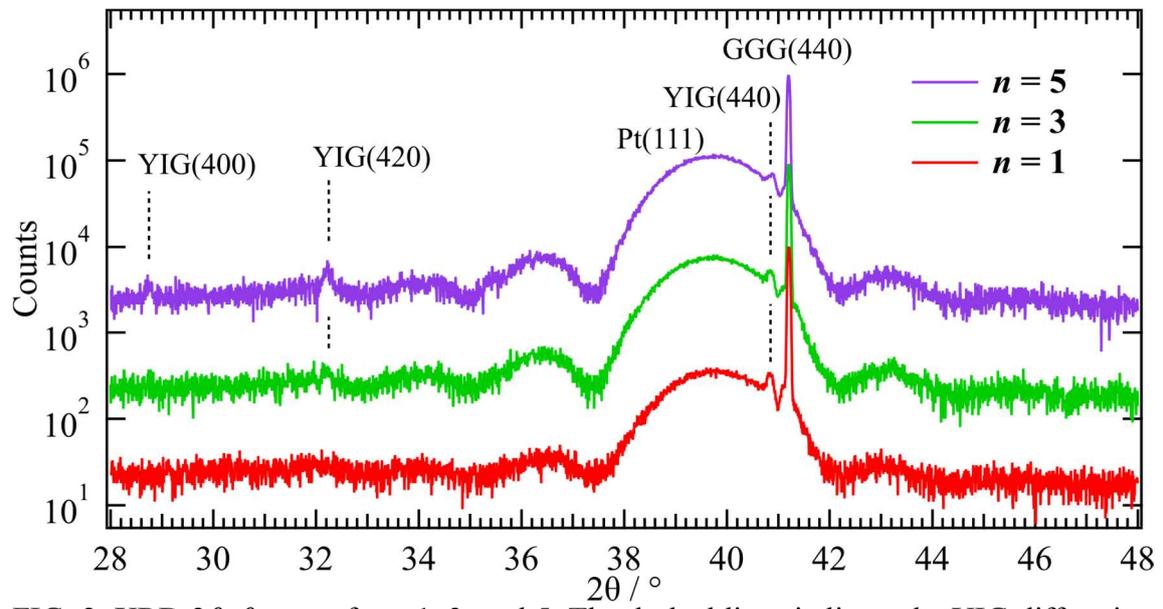

FIG. 2. XRD 2$\theta$–$\theta$ scan of $n$ = 1, 3, and 5. The dashed lines indicate the YIG diffraction peaks. The spectra of $n$ = 3 and 5 are shifted vertically for clarity.

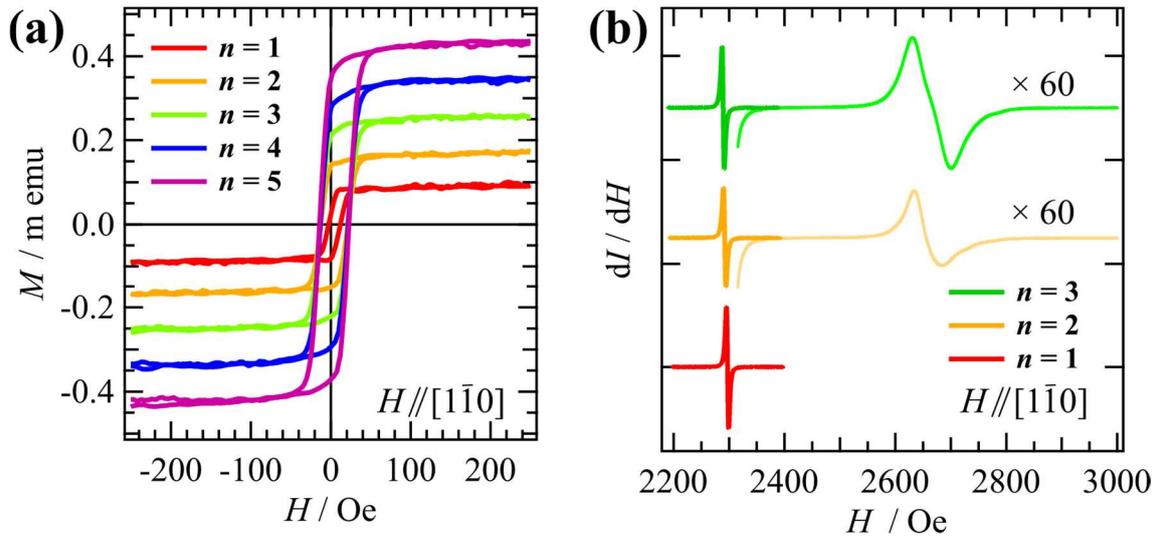

FIG. 3. (a) M–H curves of $n$ = 1 to 5. (b) Derivative FMR absorption spectra of $n$ = 1 to 3. The absorption signals of $n$ = 2 and 3 at a higher $H$ of ~2660 Oe are magnified by multiplying 60 for clarity.

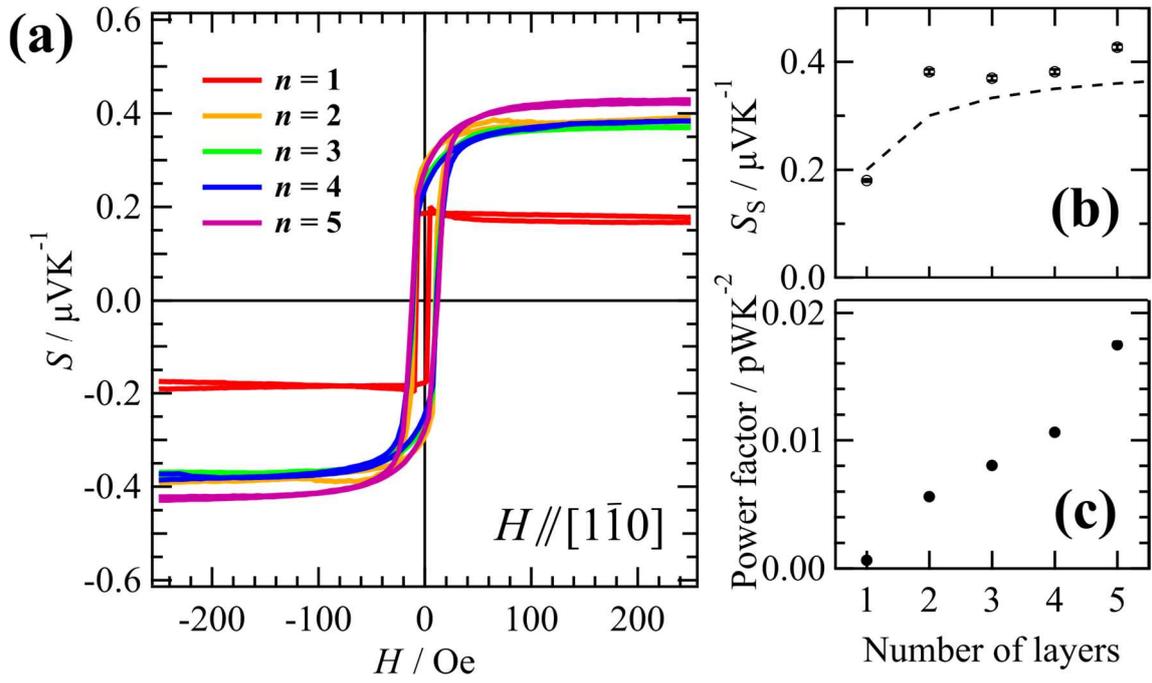

FIG. 4. (a) LSSE coefficient $S$ versus $H$ of $n$ = 1 to 5. (b) Saturation LSSE coefficient $S_s$ and (c) PF versus $n$. The dashed line in (b) denotes the calculated value of $(2-1/n)$.